\documentstyle[epsfig,twocolumn,eqsecnum,aps]{revtex}

\begin{document}
\draft
\preprint{HEP/123-qed}
\title{Relaxation Processes in Administered-Rate Pricing}

\author{Raymond J.~Hawkins$^1$ and Michael R.~Arnold$^2$ }
\address{$^1$Bear, Stearns Securities Corporation,
220 Bush Street, Suite 845, San Francisco, CA  94104 \\
$^2$Deloitte \& Touche, Capital Markets Group, 50 Fremont Street,
San Francisco, CA  94105}
\date{\today}
\maketitle
\begin{abstract}
We show how a novel application of the theory of anelasticity
unifies the observed dynamics and proposed models of
administered-rate products. This theory yields a straightforward
approach to rate model construction that we illustrate by
simulating the observed relaxation dynamics of two administered
rate products. We also demonstrate how the use of this formalism
leads to a natural definition of market friction.
\end{abstract}

\narrowtext

\section{Introduction}
Administered-rate products represent a substantial fraction of the
liabilities of banks and savings and loans\footnote{Mays
\cite{mays96} notes that nonmaturity deposits - a class of
administered-rate products - comprise ``42\% of total bank
liabilities and over 25\% of savings and loan (S\&L) liabilities
as of December 1995".} and the ability to describe the response of
these products to changes in market rates is of critical
importance for interest-rate risk management.\footnote{As pointed
out by Mays \cite{mays96} and O'Brien {\it et
al.}~\cite{obrienetal94} the present value $PV$ of
administered-rate nonmaturity deposits can be estimated using the
equation
\begin{equation}
PV = D_o - \sum_{t=1}^{\infty} { {D_{t-1} \left ( r_t^{(m)} -
r_t^{(d)} - c \right )} \over {\left ( 1 + r_t^{(m)}\right )^t} }
\; ,
\end{equation}
where $D_t$ is the deposit balance at time $t$, $r_t^{(m)}$ is the
market rate, $r_t^{(d)}$ is the administered deposit rate, and $c$
denotes the non-interest charges associated with maintaining the
account. Clearly there is interest rate sensitivity in the present
value: explicitly in the appearance of the market rate in Eq.~1
and implicitly in both the sensitivity of the balances and the
administered deposit rate to changes in the market rate.}  Unlike
market rates (e.g.~U.S~Treasury bond) that are set in an auction
environment or contractual rates that respond instantaneously to
market rates in a prescribed manner (e.g.~home mortgage),
administered rates (e.g.~interest rates on checking and savings
accounts) are set by a committee seeking an equilibrium rate in
response to changes in market rates.  Factors that bear on this
equilibrium pricing include expected future market rates,
competitor pricing responses, and depositor's short-term and
long-term balance elasticities (i.e. propensity to change balance
levels in response to rate changes); all of which are known with
limited certainty. Balance elasticities are exceedingly difficult
to estimate with any reliability because banks usually don't
preserve much historical data and don't usually employ the
resources needed to evaluate the data that does exist.
Consequently, these committees indicate a certain amount of
inertia.

The general proclivity of pricing committees to leave well enough
alone has been found in a variety of empirical
studies\footnote{See, for example, Berger and Hannan
\cite{bergerhannan89}, Ausubel \cite{ausubel90}, Diebold and
Sharpe \cite{dieboldsharpe90}, Hannan and Berger
\cite{hannanberger91}, Hannan and Liang \cite{hannanliang90},
Newmark and Sharpe \cite{newmarksharpe92}, and Hutchison and
Pennacchi \cite{hutchisonpennacchi96}.} that have established that
administered rates are driven by market rates and that the
response of an administered rate to the market rate is not
instantaneous.  This behavior has been modeled by a number of
investigators\footnote{See, for example, Hutchison and Pennacchi
\cite{hutchisonpennacchi96}, Selvaggio \cite{selvaggio96},
O'Brien, Orphanides, and Small \cite{obrienetal94}, The Office of
Thrift Supervision \cite{ots94}, Jarrow and Levin
\cite{jarrowlevin96}, Newmark and Sharpe \cite{newmarksharpe92},
Mays \cite{mays96}, and O'Brien \cite{obrien00}} who have settled
largely on the use of partial adjustment models\footnote{See, for
example, Chapter 9 of Kennedy \cite{kennedy98} for a discussion of
partial adjustment models in econometrics.} to describe the
non-instantaneous response of administered rates to changes in
market rates.  While these models often adequately describe the
observed administered-rate behavior, they largely lack a
theoretical basis with which to interpret the resulting parameters
and with which to link rate policy to the rate model.  In this
paper we show that the formal assumptions upon which previous
treatments of administered-rate dynamics are based are identical
to the assumptions underlying the formal treatment of a variety of
relaxation processes in condensed-matter physics including
magnetic, dielectric and anelastic relaxations
\cite{dattagupta87}. All these physical phenomena involve
time-dependent relaxations toward newly established equilibria
that follow from a change in a driving force and can be described
in terms of linear response theory.  Since these physical
phenomena share a common mathematical description of
relaxation/response and since these physical phenomena and
administered-rate dynamics share a variety of underlying
assumptions we make the {\it ansatz} that these phenomena all
share a common mathematical description.  Given this we can move
beyond an {\it ad hoc} treatment of administered-rate dynamics and
employ the phenomenological models that have been developed to
model these physical relaxation processes to model the dynamics of
administered-rate deposits.

The theory of anelasticity provides a useful framework with which
to develop our treatment of administered-rate dynamics because of
the similarity between some of the equations that have appeared in
the literature on administered rates and the scalar representation
of anelasticity. We show in Sec.~\ref{sec:dynamics} how the theory
of anelasticity can be used to develop a hierarchy of
continuous-time models of the administered-rate response function
that include, as a subset, the partial adjustment models that have
appeared in the literature. We find that the continuous-time
models that follow from this approach lend themselves to an easy
mapping between certain aspects of rate policy and of model
structure.  We illustrate the utility of this approach in
Sec.~\ref{sec:application} by modeling the rate-response behavior
for two administered-rate products: money market accounts and time
deposits. In Sec.~\ref{sec:friction} we explore the notion of
dissipation embodied in these relaxation dynamics to develop a
formal notion of market friction.
\section{Rate Dynamics}
\label{sec:dynamics}
\subsection{Assumptions and Econometric Models}
Fundamental to essentially all descriptions of administered-rate
product rates is the notion that there exists an equilibrium
relationship between the product rate $\mathaccent"0365{r}^{(p)}$
and the market rate $r^{(m)}$ that is of the form
\begin{equation}
\mathaccent"0365{r}^{(p)} = c + J r^{(m)} \; , \label{eq:hooke}
\end{equation}
where the tilde indicates equilibrium.  The constant $c$ is often
taken to denote the costs to the bank of servicing the product.
The proportionality factor $J$ has, in the case of nonmaturity
deposits, been interpreted as the fraction of deposited monies
that Federal Reserve requirements allow to be lent\footnote{See,
for example, O'Brien {\it et al.} \cite{obrienetal94}, Jarrow and
van Deventer \cite{jarrowvandeventer98}, and Janosi {\it et
al.}\cite{janosietal99} where $J$ is taken to be one minus the
marginal reserve rate.}. While the interpretation of $J$ in terms
of a Federal Reserve requirement clearly breaks down for products
without such a requirement, Eq.~\ref{eq:hooke} remains,
nevertheless, a basic assumption of equilibrium behavior for most
administered-rate products.

The rate relationship in Eq.~\ref{eq:hooke} is characterized by
three features: (i) a unique equilibrium product rate for each
level of the market rate, (ii) instantaneous achievement of the
equilibrium response, and (iii) linearity of the response.  We
note in passing that the equilibrium rate is completely
recoverable.

The empirical dynamics of administered rates, however, demonstrate
that the equilibrium response is not achieved instantaneously and
a lagged response is observed.  To incorporate this observed lag
into the relationship between the product rate and the market rate
previous research has augmented Eq.~\ref{eq:hooke} with an {\it ad
hoc} ``partial adjustment" model of the form
\begin{equation}
\Delta r_n^{(p)} = \sum_{i=0}^N \left [ a_i r_{n-i}^{(m)} + b_i
r_{n-i+1}^{(p)}\right ] \; , \label{eq:paf}
\end{equation}
where we have introduced the time-dependent notation $r_n^{(p)}
\equiv r^{(p)}(t_n)$.  This functional for is the basis of most of
the econometric studies that have appeared in the literature to
date.  While some researchers have posited that the coefficients
in Eq.~\ref{eq:paf} depend on the direction of the change in the
market rate, O'Brien \cite{obrien00} has recently pointed out,
however, that such an assumption is not consistent with the
assumed equilibrium relationship given in Eq.~\ref{eq:hooke}.
\subsection{Anelastic Rate Dynamics}
The theory of anelasticity is a generalization of the theory of
ideal elasticity that allows for time dependence in the response
of a material to an applied stress.  Like previous treatments of
administered-rate dynamics it assumes the existence of a unique
equilibrium relationship between stress and strain known as
Hooke's law. The equilibrium relationship given in
Eq.~\ref{eq:hooke} is, in fact, identical to the scalar version of
Hooke's law of ideal elasticity $ \epsilon = J \sigma $ with the
product rate playing the role\footnote{While the notions of
strain, stress, and force are also used as metaphors in economics
and finance, our use of these metaphors used here is to motivate
the comparison of notation.} of strain $\epsilon$ and the market
rate playing the role of stress $\sigma$.\footnote{A more complete
correspondence with scalar elasticity can be achieved by positing
that market rates change in response to market stress
$\sigma^{(m)}$ induced by market forces. To the extent that market
rates respond to market forces in an essentially instantaneous
manner, we can write $r^{(m)} = (K/J) \sigma^{(m)}$ from which it
follows that $\mathaccent"0365{r}^{(p)} = c + K \sigma^{(m)}$:
Hooke's law relating the product rate to the market stress via the
compliance $K$.}  In this context the constant $c$ can be
interpreted as the contribution to the product rate driven by a
different "stress": the cost to the bank of maintaining the
account.\footnote{We thank Leif Wennerberg for this observation.}

The differential dynamics of anelasticity are not obtained through
an direct application of lagged variables via Eq.~\ref{eq:paf}
but, rather, by noting that the assumption of linearity implies a
general market-product rate relationship of the form
\begin{eqnarray}
& \left ( f_0 + f_1 \frac{d}{dt} + f_2 \frac{d^2}{dt^2} + \cdots
\right ) & r^{(p)}  = \nonumber \\ & \left ( g_0 + g_1
\frac{d}{dt} + g_2 \frac{d^2}{dt^2} \cdots \right ) & r^{(m)} \; .
\label{eq:general}
\end{eqnarray}
While the econometric application of this equation, like
Eq.~\ref{eq:paf}, requires an analysis of the number of terms
needed to describe the observed dynamics, the use of
Eq.~\ref{eq:general} enables an economic interpretation of these
terms and the coefficients. In practice a wide range of relaxation
dynamics have been found to be described well by the comparatively
simple differential relationship\footnote{Indeed, this
relationship is so ubiquitous that the resulting anelastic system
is referred to as the ``standard anelastic solid"
\cite{nowickberry72}.  Higher order differential equations can be
used to treat more complex relaxation processes.  Nowick and Berry
\cite{nowickberry72} show, however, that these higher-order
differential processes can be represented as a linear combination
of the standard anelastic solid.  The interest rate dynamics
discussed in this paper are described quite adequately with the
standard anelastic differential equation given above.}
\begin{equation}
\frac{d r^{(p)}}{d t} + \eta \left ( r^{(p)} - c \right ) = J_U
\frac{d r^{(m)}}{d t} + \eta J_R r^{(m)} \; , \label{eq:sas}
\end{equation}
where $\eta$ denotes the rate at which the product rate relaxes to
the equilibrium level, $J_U$ denotes that fraction of the response
that occurs instantaneously, and $J_R$ denotes the ultimate extent
of the response function ($= J(t=\infty)$).  The change in the
product rate with respect to time is, in this case, a function of
the current product rate, the current market rate, and the change
in the market rate with respect to time.

\begin{figure}
\begin{center}
\epsfig{file=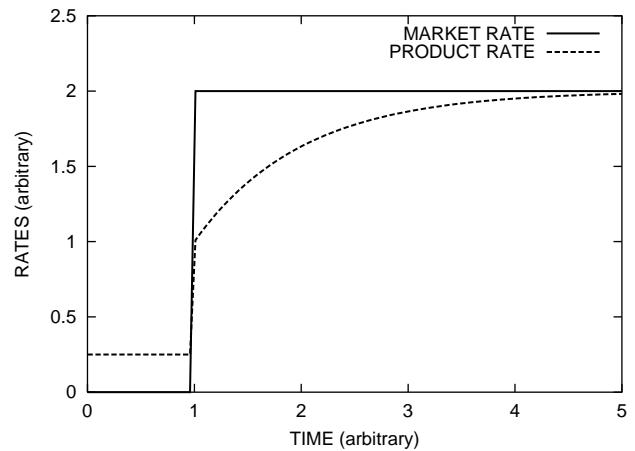}
\end{center}\caption{The response of the product rate to a step change in the
market rate given by Eq.~\ref{eq:sas}.} \label{fig:response}
\end{figure}

Some intuition for the interpretation of this differential
product-market rate relationship can be obtained for the case of a
simple market rate shock. Given a sudden change in the market rate
that is subsequently held constant at $r^{(m)}$ and the
equilibrium relationship given by Eq.~\ref{eq:hooke},
Eq.~\ref{eq:sas} can be integrated to yield the time-dependent
product rate
\begin{equation}
r^{(p)}(t) =  c + \left ( J_U + \delta J \left [ 1 - e^{-\eta t}
\right ]
 \right ) r^{(m)}\; ,
\end{equation}
where $\delta J \equiv \left [ J_R - J_U \right ] $, whence
\begin{equation}
J(t) = J_U + \delta J \left [ 1 - e^{-\eta t} \right ] \; ;
\label{eq:sasj}
\end{equation}
illustrating the decomposition of the response $J$ into an
instantaneous contribution $J_U$ and a time-dependent portion
proportional to $\delta J$ mentioned above.  The product-rate
response to this step change in the market rate is illustrated in
Fig.~\ref{fig:response} where we show the response of the product
rate to a step change in the market rate with $c=0.25$,
$J_U=0.375$, $\delta J = 0.5$, and $\eta = 1.0$. The product rate
tracks the market rate instantaneously over a range defined by
$J_U$; in this case to 1.0.  The product rate then relaxes to
equilibrium with the market rate (in this example
$\mathaccent"0365{r}^{(p)} = 0.25 + 0.875 r^{(m)}$). Varying
$J_U$, and $J_R$ (or, equivalently $\delta J$) one can span the
range of responses from being completely instantaneous $J_U = J_R
> 0$ to being completely time dependent $J_U = 0$.

\subsection{Anelastic Partial Adjustment Models} A variety of partial adjustment
models can be developed for the differential relationship given
above by discretizing\footnote{See, for example, Chapter 1 of
Koonin \cite{koonin86} or Section 16.7 of Press {\it et
al.}~\cite{pressetal94}.} the integral form of $dr^{(p)}/dt$ given
in Eqs.~\ref{eq:sas}
\begin{equation}
r^{(p)}(t) = r_n^{(p)} + \int_{t_n}^t
f(r^{(p)}(\tau),r^{(m)}(\tau),c) d\tau \; ,
\end{equation}
which is known to yield
\begin{equation}
r_{n+1}^{(p)} = r_n^{(p)} + h \sum_{i=0}^N \beta_i f_{n+1-i} \; ,
\end{equation}
where $f$ represents all terms in Eq.~\ref{eq:sas} except for
$dr^{(p)}/dt$ and $h$ is the time step. This expression is
essentially identical to the partial adjustment formula given in
Eq.~\ref{eq:paf} above. Using standard discretization techniques
one can derive several partial adjustment models from
Eq.~\ref{eq:sas} appropriate for more complex behaviors of the
driving market rate. We now explore some common discretizations of
our standard anelastic system that resemble closely the partial
adjustment formulae that have appeared in the literature.

Rewriting Eq.~\ref{eq:sas} in terms of the variable $y \equiv
r^{(p)} - J_U r^{(m)}$, we begin with the forward-difference Euler
equation $y_{n+1} = y_n + f_n$ from which it follows that
\begin{eqnarray}
r_{n+1}^{(p)} & = & \eta c + r_{n}^{(p)} + J_U \left (
r_{n+1}^{(m)} - r_{n}^{(m)} \right )  \nonumber \\ & + & \eta
\left ( J_R r_n^{(m)}- r_n^{(p)} \right ) \label{eq:pam1}
\end{eqnarray}
We note that when $\eta J_R = J_U$ this relationship becomes
\begin{equation}
r_{n+1}^{(p)} = \eta c + \left ( 1 - \eta \right ) r_{n}^{(p)} +
J_U r_{n+1}^{(m)} \label{eq:pam2}
\end{equation}
which is identical in structure to the simplest partial adjustment
model discussed by Mays \cite{mays96}.

While the Euler discretization of Eq.~\ref{eq:sas} yields a
popular partial adjustment expression, partial adjustment models
with more temporal lags do exist in the literature on
administered-rate products.  More temporal lags can be introduced
in two different ways. First, if the deliberations of the pricing
committee are known to correspond to Eq.~\ref{eq:sas} (i.e. the
product rate is based on considerations of the level and change of
the rates) then partial adjustment models can be obtained using
different discretization techniques. Alternatively, greater
temporal lags follow naturally if the deliberations of the
committee include the change in the slope of the market and
product rates as a function of time.  We examine each in turn.

Partial adjustment models based on Eq.~\ref{eq:sas} with greater
temporal lags can be obtained through higher-order discretizations
such as the Adams-Bashford methods. The Adams-Bashford two-step
method is given by $y_{n+1} = y_{n} + \left [ \frac{3}{2} f_n -
\frac{1}{2}f_{n-1} \right ] $. Applying this to Eq.~\ref{eq:sas}
yields
\begin{eqnarray}
\label{eq:pam3} r_{n+1}^{(p)} & = & \eta c + r_n^{(p)}+ J_U \left
(  r_{n+1}^{(m)} - r_{n}^{(m)} \right ) \\
 & + &
\frac{3}{2} \eta \left ( J_R r_{n}^{(m)} - r_{n}^{(p)} \right ) -
\frac{1}{2} \eta \left (  J_R r_{n-1}^{(m)} - r_{n-1}^{(p)} \right
) \; \nonumber ,
\end{eqnarray}
which shares many structural aspects with the partial adjustment
model developed by the Office of Thrift Supervision (OTS)
\cite{ots94}.  Further temporal lags can be included by applying
the Adams-Bashford three-step method $y_{n+1} = y_n + \left [
\frac{3}{2} f_n - \frac{1}{2} f_{n-1} \right ]$ to
Eq.~\ref{eq:sas}:
\begin{eqnarray}
\label{eq:pam4} r_{n+1}^{(p)} & = & \eta c + r_n^{(p)} + J_U \left
(  r_{n+1}^{(m)} - r_{n}^{(m)} \right ) \\ & + & \frac{23}{12}
\eta \left (  J_R r_{n}^{(m)} - r_{n}^{(p)} \right ) -
\frac{16}{12} \eta \left (  J_R r_{n-1}^{(m)} - r_{n-1}^{(p)}
\right ) \nonumber \\ & + & \frac{5}{12}  \eta \left (  J_R
r_{n-2}^{(m)} - r_{n-2}^{(p)} \right ) \; \nonumber .
\end{eqnarray}

If the product rate is thought also to be a function of the change
in the slope of the market and product rates as a function of
time, it is likely that the dynamics are better represented by the
relationship involving two relaxations
\begin{eqnarray}
&& \frac{d^2 r^{(p)}}{d t^2} + \left ( \eta^{(1)} + \eta^{(2)}
\right ) \frac{d r^{(p)}}{d t} +  \eta^{(1)}\eta^{(2)} \left (
r^{(p)} - c \right ) \nonumber \\  & = & J_U \frac{d^2 r^{(m)}}{d
t^2} \nonumber \\ & + & \left [\delta J^{(1)}\eta^{(1)} + \delta
J^{(2)}\eta^{(2)}
  + \left (\eta^{(1)} + \eta^{(2)} \right )J_U \right ] \frac{d
r^{(m)}}{d t} \nonumber \\
 & + & \eta^{(1)}\eta^{(2)} \left( \delta J^{(1)} + \delta J^{(2)} +
J_U \right )r^{(m)} \; , \label{eq:sas2}
\end{eqnarray}
which is simply Eq.~\ref{eq:general} with up to second-order
derivatives included and where $\eta^{(i)}$ and $\delta J^{(i)}$
correspond to the $i^{th}$ relaxation. While somewhat more
formidable than Eq.~\ref{eq:sas}, given this choice of terms and
coefficients the associated response function is known
\cite{nowickberry72} to be a simple sum of the responses resulting
in a generalization of Eq.~\ref{eq:sasj} to
\begin{equation}
J(t) = J_U + \sum_{i=1}^2 \delta J^{(i)} \left [ 1 -
e^{-\eta^{(i)} t} \right ] \; , \label{eq:sasj2}
\end{equation}
where we see that the response function is now contains two
relaxation response terms in addition to the instantaneous
response.

The partial adjustment model that follows from an Euler
discretization of Eq.~\ref{eq:sas2} yields, via the second
derivatives, a function of the form
\begin{equation}
r_{n+1}^{(p)} = \eta c + r_n^{(p)} + h(r_n^{(p)},\;
r_{n-1}^{(p)},\;r_{n+1}^{(m)},\;r_n^{(m)},\;r_{n-1}^{(m)}) \; .
\end{equation}
While this expression contains the same number of temporal lags as
Eq.~\ref{eq:pam3} and has a structure similar to the of the OTS
model \cite{ots94}, it has 2 more degrees of freedom than
Eq.~\ref{eq:pam3} due to the more complex relaxation process.

The anelastic partial adjustment models differ from the partial
adjustment models that have appeared in the literature in two
important ways. First, number of free parameters is determined by
the nature of the differential rate relationship, while the number
of terms is determined independently by the nature of the
discretization of the differential rate relationship.  Second, the
decoupling of the number of terms from the number of free
parameters in the partial adjustment model allows the partial
adjustment model to better reflect the nature of the how pricing
committees adjust rates in response to market rates. While the
differential rate relationship provides a phenomenological
representation of the observed lag in the product-rate adjustment,
the discrete form of the model can be expressed so as to reflect
the number of previous time periods included in the deliberations
of the pricing committee.  Thus, the anelastic approach provides a
convenient way to avoid overspecification of the relationship
between the product and market rates while simultaneously
providing the flexibility needed to properly represent the data
used in rate policy decisions.
\subsection{Simulation using Boltzmann Superposition}
Having demonstrated that the functional form of the partial
adjustment models that have appeared in the literature can be
recovered from the anelastic formalism using standard
discretization techniques, we note in passing that there is a far
simpler approach to the modeling of these rate dynamics that may
be of use in future work.  As pointed out by Nowick and Berry
\cite{nowickberry72} manipulation (and discretization) of the
differential product-market rate equations becomes increasingly
complex as the order of these equations increase.  An
appropriately chosen manner of increasing the order of the
differential equation (employed, for example, in
Eq.~\ref{eq:sas2}) allows us to write down the response function
(cf. Eqs.~\ref{eq:sasj} and \ref{eq:sasj2}) directly as
\begin{equation}
J(t) = J_U + \sum_{i=1}^N \delta J^{(i)} \left [ 1 -
e^{-\eta^{(i)} t} \right ] \; , \label{eq:boltzmann}
\end{equation}
where $N$ represents both the order of the differential terms
included in Eq.~\ref{eq:general} and the number of relaxation
terms.  Then, as a consequence of the linearity of the system -
aka the Boltzmann superposition principle, we can, given a history
of market rate changes $r_i^{(m)}$ applied at successively
increasing times $\tau_1,\; \tau_2,\; \ldots,\; \tau_M$ write the
product rate as
\begin{equation}
r^{(p)}(t) - c = \sum_{i=1}^M r_i^{(m)} J(t-\tau_i) \; .
\end{equation}
These simple relations provide a straightforward calculation of
the product rate in response to changes in the market rate with a
response function that is specified easily in terms of 2 types of
coefficients: a relaxation rate and the fraction of the response
corresponding to that rate.

While by no means an exhaustive collection of the models that can
be generated from Eqs.~\ref{eq:hooke} and \ref{eq:general},
Eqs.~\ref{eq:pam1}, \ref{eq:pam2}, \ref{eq:pam3}, \ref{eq:pam4},
and \ref{eq:boltzmann} illustrate, nevertheless, the rich variety
of partial adjustment models the follow from a single linear
differential relationship between the product and market rates
that, in turn, follows from an anelastic interpretation of the
relationship between the product and market rates. We now apply
the anelastic approach to the description of the observed dynamics
of administered rates.

\section{Anelastic Relaxations and Observed Product Rate Dynamics}
\label{sec:application}
We have used the anelastic approach
described above to model the dynamics of the Cash Maximizer$^{\rm
TM}$ (CMX) interest rates \cite{bofa99} and of the rates for
retail certificates of deposit (CDs).  The Cash Maximizer$^{\rm
TM}$ account is a money market deposit account that requires a
minimum of USD 2,500 to open and to avoid service
charges\footnote{In July 1986 Bank of America introduced USD
25,000 and USD 100,000 tiers to this account. While the
introduction of these tiers did introduce some additional pricing
constraints (e.g. higher-minimum tiers have rates greater than or
equal to lower-minimum tiers) the dynamics of each tier is
otherwise considered to be independent of the other tiers.}.
Unlike the CMX product that has no defined maturity, the retail CD
used in this study has a 3-month maturity\footnote{Retail
customers of Bank of America could, during the period to be
analyzed, select almost any maturity less than 7 years.  Most
customers chose the conventional maturities of 3 months, 6 months,
or annual increments out to 7 years.}. The CD that we shall
analyze below has a face value of USD 2,500. The CD and CMX rates
are set by committee.

The market rate driving changes in the CMX rates is taken to be
that rate most closely matching a matched rate along the bank's
cost-of-funds curve.  Since the marginal cost of funds for Bank of
America during this period is best reflected by the London
Interbank Offer Rate (LIBOR) for short maturities, we have taken
3-month LIBOR to be the market rate.

\begin{figure}
\begin{center}
\epsfig{file=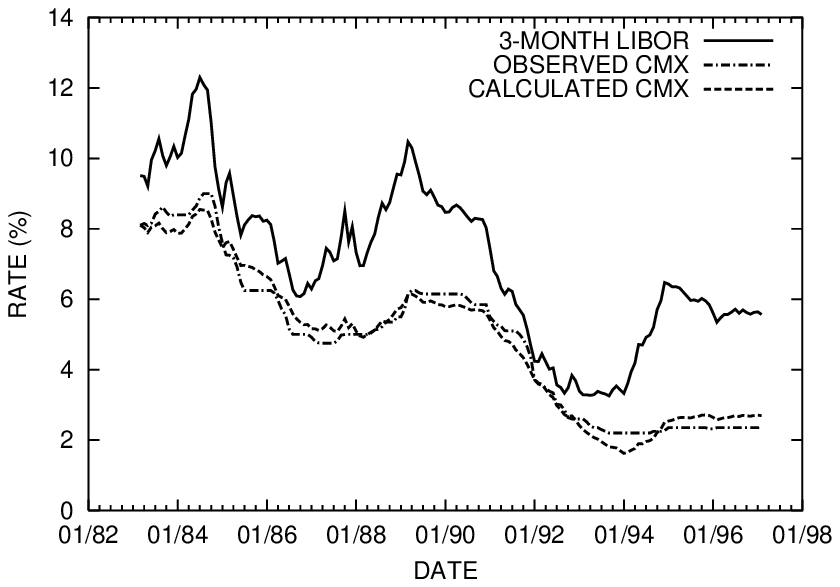}
\end{center}\caption{The market rate (3-month LIBOR), observed CMX rate, and
calculated CMX rate as a function of time.} \label{fig:cmx}
\end{figure}

The month-end CMX rates are shown together with 3-month LIBOR for
the period March 1983 - February 1997 in Fig.~\ref{fig:cmx}.
Comparing these rates we see that the CMX rate is always less than
3-month LIBOR and that the CMX rate roughly tracks the movements
of 3-month LIBOR in a largely attenuated and somewhat lagged
manner. We fit Eqs.~\ref{eq:pam1}, \ref{eq:pam3}, and
\ref{eq:pam4} to these data using the Generalized Reduced Gradient
(GRG2) nonlinear optimization solver in Microsoft Excel$^{\rm
TM}$. The coefficients resulting from these fits are given in
Table~\ref{tab:cmx}. The fit of Eq. \ref{eq:pam1} is shown
together with the CMX rates and the 3-month LIBOR rates in Fig.~2:
the other fits are not shown as differences among them cannot be
resolved by eye on this scale.

\begin{figure}
\begin{center}
\epsfig{file=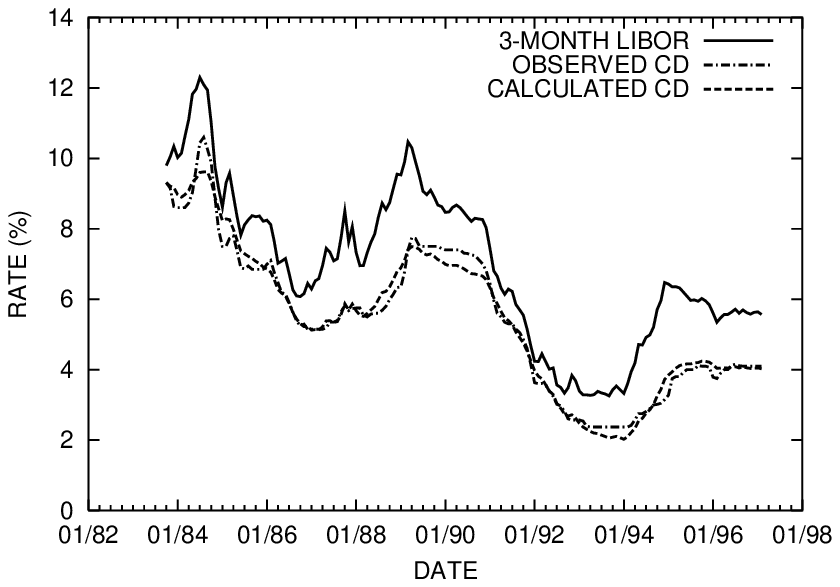}
\end{center}\caption{The market rate (3-month LIBOR), observed CD rate, and
calculated CD rate as a function of time.} \label{fig:cd}
\end{figure}

An identical analysis was performed on the month-end CD rates
shown, together with 3-month LIBOR for the period October 1983 -
February 1997, in Fig.~\ref{fig:cd}.  The results of the fitting
described above are shown in Table \ref{tab:cd}.  The fit of Eq.
\ref{eq:pam1} is shown together with the CD rates and 3-month
LIBOR in Fig.~\ref{fig:cd}. As in the case of Fig.~\ref{fig:cmx},
differences among the various fits cannot be meaningfully resolved
by eye on this scale.

We see in Figs.~\ref{fig:cmx} and \ref{fig:cd} that modeling
changes in these product (CMX, CD) rates as an anelastic response
to a market rate changes results in an expression (e.g.
Eq.~\ref{eq:pam1}) that can track the observed product rate quite
well.  This is remarkable given that these two products have quite
different maturity assumptions: the CD has a fixed maturity while
the CMX product has no maturity. These results, together with the
observation made above that economic assumptions concerning the
relationship between the market and product rates are the same as
those of an anelastic process, provide compelling evidence that
these product rates respond to market rates as if via anelastic
relaxations.
\section{Relaxations and Market Friction}
\label{sec:friction} In a mechanical system the time dependent
stress-strain behavior is "an external manifestation of internal
relaxation behavior that arises from a coupling between stress and
strain through internal variables that change to new equilibrium
values only through kinetic processes such as diffusion"
\cite{nowickberry72}. Similarly, time dependent
market-administered rate behavior is an external manifestation of
internal relaxation behavior that arises from a coupling between
the market and product rates through internal variables such as
competitor pricing responses and depositors' balance elasticities
that change to new equilibrium values only after the passage of
time.  In both mechanical and market systems this temporal lag in
response to an applied force is a manifestation of friction.

Our identification of administered-rate dynamics as relaxations
also provides a way of quantifying market friction. An expression
for this dissipation - also known as ``internal friction" in the
anelasticity literature - can be obtained by considering the case
of a periodic market scalar stress $\sigma^{(m)}(t)$ due to a
periodic market force
\begin{equation}
\sigma^{(m)}(t) = \sigma^{(m)}(0) e^{i \omega t} \; ,
\end{equation}
where $\sigma^{(m)}(0)$ is the market stress at time $t = 0$, $i =
\sqrt{-1}$, and $\omega$ is the cyclic frequency of the market
stress.  The product rate will track the market force (and market
rate, due to linearity as discussed above) with a lag that can be
represented by a loss angle $\phi$:
\begin{equation}
r^{(p)}(t) = r^{(p)}(0) e^{i (\omega t - \phi)} \; .
\end{equation}
These expressions for the market and product rates imply a
frequency dependent proportionality factor $J(\omega)$ (the
Fourier transform of $J(t)$) that is complex $J(\omega) =
J_1(\omega) - i J_2(\omega)$ and a loss angle related to the
components of $J(\omega)$ by $\tan (\phi) =
J_2(\omega)/J_1(\omega)$.

\begin{figure}
\begin{center}
\epsfig{file=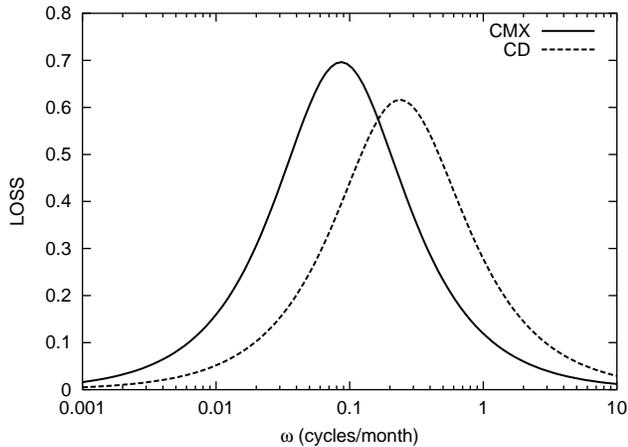}
\end{center}
\caption{The loss curves for
CMX and CD rates.} \label{fig:loss}
\end{figure}

The isomorphism between anelasticity and administered rates
implies the existence of a state variable - a market equivalent of
energy - at any phase in the market cycle given by $\int
\sigma^{(m)} dr^{(p)} \propto \int r^{(m)} dr^{(p)}$ taken between
the start of the cycle up to the point of interest. The market
energy dissipated in a full market cycle is
\begin{equation}
\Delta U = \oint \sigma^{(m)} dr^{(p)} \propto \pi J_2 \left [
r^{(m)}(0) \right ]^2 \; ,
\end{equation}
and the stored market energy is
\begin{equation}
U = \int_{\omega t = 0}^{\pi / 2} \sigma^{(m)} dr^{(p)} \propto
\frac{1}{2} J_1 \left [ r^{(m)}(0) \right ]^2 \; .
\end{equation}
The ratio of these terms - the fractional market energy dissipated
in a full market cycle - is related to the loss angle $\phi$ by
$\Delta U/U = 2 \pi \tan (\phi)$ which, for the administered rates
described by Eq. \ref{eq:sas} yields
\begin{equation}
\tan \phi = \delta J \frac{\omega / \eta}{J_R + J_U \omega^2 /
\eta^2}. \label{eq:friction}
\end{equation}
thus we see that the existence of an anelastic response ($\delta J
\neq 0$) in a market system implies a dissipation of market energy
and a formal definition of market friction.  As the existence of
this loss angle is due to the gathering and processing of
information needed to reestablish equilibrium - ${r}^{(p)}
\rightarrow \mathaccent"0365{r}^{(p)}$ - it follows that the loss
angle is also a measure of relative efficiency discussed by Farmer
and Lo \cite{farmerlo99}.  This is illustrated in
Fig.~\ref{fig:loss} where we see the loss angle as a function of
cyclic frequency $\omega$ for both the CMX and CD rates.  In the
limit $\omega \rightarrow 0$ the market cycle become so long that
any finite relaxation rate is, on the time scale of a market
cycle, instantaneous and the rate systems behave as if there is no
relaxation with $J(t) = J_R$.  In the limit $\omega \rightarrow
\infty$ the market cycle becomes so short that the only component
of the system that responds to the market force is the
instantaneous portion $J_U$. Between these extremes we see the
loss due to the relaxation behavior. For $\omega < 0.16$ cycles
per month (cycles greater than about 6 months) the CMX rate loss
is greater than that of the CD.  This is consistent with our
expectation that the rate system that relaxes faster requires less
market energy to achieve equilibrium and is, therefore, less lossy
and more efficient.  For $\omega > 0.16$ (cycles less that about 6
months) the CD rate has a greater fractional loss per cycle.  This
perhaps unexpected result follows from the relative relaxation
rates of these two systems.  As the cyclic frequency increases the
relaxation component begins to "freeze out" and the system behaves
in an increasingly elastic manner with $J(t) \rightarrow J_U$.
Since the CD rate relaxes faster than the CMX rate, the relaxation
component of the CMX rate "freezes out" first making it less lossy
than the CD rate in this frequency range.
\section{Discussion and Summary}
Administered rates are unique in that they are set by a group of
individuals attempting to maximize profits in the face of market
forces.  As the future direction of market forces, commonly
measured by market rates, is unknown and committee decisions
exhibit a degree of inertia, equilibrium between the market force
and administered rate is achieved only after the passage of a
certain amount of time. Historically this process has been
formally expressed by an assumed linear equilibrium relationship
and an {\it ad hoc} partial adjustment model to describe the
change of the administered rate in response to a change in the
market rate. Common to most previous treatments of
administered-rate dynamics are the postulates that (i) for every
market rate there is a unique equilibrium rate, and vice versa,
(ii) the equilibrium response is achieved only after the passage
of sufficient time, and (iii) the market-administered rate
relationship is linear.  A contribution of this paper is the
observation that these postulates also form the basis for a
well-developed theory of relaxation processes in the physical
sciences:  indeed, these postulates are paraphrased directly from
the introduction to anelasticity presented by Nowick and Berry
\cite{nowickberry72}.  We have examined this market system and
found that the assumed equilibrium rate relationship corresponds
to Hooke's law of elasticity and that the relaxation dynamics of
administered rates are quite similar to anelastic relaxations.
Developing this isomorphism we demonstrated that the basic
structure of popular {\it ad hoc} partial adjustment models could
be reproduced easily using standard techniques for discretizing
the simplest anelastic differential relationship between the
administered rate and market rate. We applied these models to the
observed interest rate dynamics of a Cash Maximizer$^{\rm TM}$
account and a certificate of deposit and found that, in spite of
significant differences in the maturity features of these
products, their dynamics are described well as anelastic
relaxations.  Finally we found that the anelastic description of
these dynamics provides a natural definition of market friction as
the realization of internal friction or dissipation in this market
system.

\acknowledgments

We thank Leif Wennerberg for criticisms of a previous draft that
significantly improved this paper, David Anderson for help with
the data used in this study and Prof.~Daniel N.~Beshers for
enlightening conversations.  Neither Bear, Stearns Securities
Corporation or Deloitte \& Touche are responsible for any
statements or conclusions herein; and no opinions, theories, or
techniques represented herein in any way represent the position of
Bear, Stearns Securities Corporation or of Deloitte \& Touche.

\begin{table}
\caption{Fitted parameters and the sum of the squared errors (SSE)
for the (i) Euler forward difference (Euler FD), (ii)
Adams-Bashford two-step (Adams-Bashford 2), and (iii)
Adams-Bashford three-step (Adams-Bashford 3) discretized forms of
the anelastic model for the CMX rate.} \label{tab:cmx}
\begin{tabular}{cccc}
     & Euler FD & Adams-Bashford 2 & Adams-Bashford 3 \\ \tableline
$c$    & -2.8544    & -2.7653  &  -2.7006 \\ $J_U$  &  0.2933    &
0.2690  &   0.2633 \\ $J_R$  &  1.0060    &  0.9938  &   0.9854 \\
$\eta$ &  0.0430    &  0.0452  &   0.0464 \\ $SSE$  & 19.1702    &
18.5605  &  18.3177 \\
\end{tabular}
\end{table}

\begin{table}
\caption{Fitted parameters and the sum of the squared errors (SSE)
for the (i) Euler forward difference (Euler FD), (ii)
Adams-Bashford two-step (Adams-Bashford 2), and (iii)
Adams-Bashford three-step (Adams-Bashford 3) discretized forms of
the anelastic model for the CD rate.} \label{tab:cd}
\begin{tabular}{cccc}
     & Euler FD & Adams-Bashford 2 & Adams-Bashford 3 \\
     \tableline $c$    & -1.1967 &  -1.2097 & -1.3271 \\ $J_U$  &
0.3255 & 0.2932 &  0.3331 \\ $J_R$  &  0.9289 &   0.9307 &  0.9493
\\ $\eta$ & 0.1275 &   0.1332 &  0.1228 \\ $SSE$  & 13.7333 &
13.3897 & 13.1701 \\
\end{tabular}
\end{table}

\end{document}